# Analysis and perturbation of degree correlation in complex networks

Ju Xiang[1,2,3], Ke Hu[4(*)], Yan Zhang[5(*)], Tao Hu[6] and Jian-Ming Li[1,2]

[1]*Neuroscience Research Center, Changsha Medical University, Changsha 410219, China;*
[2]*Department of Anatomy, Histology and Embryology, Changsha Medical University, Changsha 410219, China*
[3]*Department of Basic Medical Sciences, Changsha Medical University, Changsha 410219, China*
[4]*Department of Physics, Xiangtan University, Xiangtan 411105, Hunan, China*
[5]*Department of Computer Science, Changsha Medical University, Changsha 410219, China*
[6]*College of Science, QiLu University of Technology, Jinan 250353, Shandong, China*

* Corresponding author. Email: huke1998@aliyun.com, xiang.ju@foxmail.com, zhangyancsmu@foxmail.com

**Abstract** – Degree correlation is an important topological property common to many real-world networks such as such as the protein-protein interactions and the metabolic networks. In the letter, the statistical measures for characterizing the degree correlation in networks are investigated analytically. We give an exact proof of the consistency for the statistical measures, reveal the general linear relation in the degree correlation, which provide a simple and interesting perspective on the analysis of the degree correlation in complex networks. By using the general linear analysis, we investigate the perturbation of the degree correlation in complex networks caused by the simple structural variation such as the "rich club". The results show that the assortativity of homogeneous networks such as the Erdös-Rényi graphs is easily to be affected strongly by the simple structural changes, while it has only slight variation for heterogeneous networks with broad degree distribution such as the scale-free networks. Clearly, the homogeneous networks are more sensitive for the perturbation than the heterogeneous networks.

PACS: 89.75.Hc - Networks and genealogical trees
PACS: 89.75.Fb - Structures and organization in complex systems

**Introduction. -** Complex networks provide a useful tool for investigating the topological structure and statistical properties of complex systems with networked structures [1]. These networked systems have been found to possess many common topological properties. For example, many real-world networks such as the protein-protein interactions, the metabolic networks and the Internet exhibit the existence of the nontrivial correlation between degrees of nodes connected by edges [2-5]. Empirical studies show that almost all social networks display the property that high- or low-degree nodes tend to connect to other nodes with similar degrees, which is referred to as "assortative mixing". In biological and technological networks, high-degree nodes often preferably connect to low-degree nodes, which is referred to as "dissassortative mixing". The degree correlation has important influence on the topological properties of networks and may impact related problems on networks such as stability [6], the robustness of networks against attacks [7], the network controllability [8], the traffic dynamics on networks [9, 10], the network synchronization [11-13], the spreading of information or infections and other dynamic processes [7, 13-24].

In order to characterize and understand such preference of connections in complex networks, many statistical measures and network models have been introduced and investigated [2, 3, 25-36]. For example, the average nearest neighbors' degree of nodes (ANND) [3] and the degree correlation coefficient (DCC) [2] are the classical measures for characterizing the degree correlation in networks. ANND could give the intuitional information of the connection preference in degree, and DCC is helpful for the quantitative comparison of the different networks. As an integral characterization of the degree correlation, DCC may miss some details of the degree correlation [37]. The generalized degree correlation coefficient proposed by Valdez et al [15] is a useful extension of DCC, which can help compare the difference of networks with the same DCC and exhibit the correlations at different levels. Generally, it may be more appropriate that the combinations of the measures are used for the analysis of the correlations in networks.

In the letter, we will study analytically the classical statistical measures for characterizing the degree correlation in complex networks, and give some interesting results, including the exact proof of the consistency of the measures and the general linear relation in the degree correlation, which provides a simple and interesting perspective on the analysis of the degree correlation in networks. In terms of the general linear relation in the degree correlation, we analyze the perturbation of the degree correlation caused by the addition of few nodes and the "rich club".

**Analysis of degree correlation in complex networks. -** The degree correlation between two nodes connected by edges can be naturally characterized by the degree-degree joint probability $e_{jk}$, the probability that one of the two ends of a randomly selected edge in a network will has node of degree $j$ and another will has node of degree $k$. This quantity is a symmetric matrix in undirected network ($e_{jk} = e_{kj}$), and have $\sum_{jk} e_{jk} = 1$ and $\sum_j e_{jk} = q_k$, where $q_k = kp_k / \sum_j jp_j$ is the probability that the end of a randomly chosen edge in a network has a node of degree $k$, while $p_k$ is the degree distribution of the network—the probability that a randomly chosen node in the network has degree $k$ (the degree of a node is the number of other nodes to which it connects in the network). If $e_{jk}$ takes the value of $q_j q_k$ in a network, the network is usually considered to have no correlation of degrees. While most of real-world networks always exhibit an obvious deviation from the value, meaning the existence of degree correlation in the networks. However, the degree-degree joint probability is easily affected by statistical fluctuations in finite networks, and it is rather difficult to identify the tendency of degree correlation in the network by the quantity. Therefore, other statistical measures may be more convenient and efficient.

*Statistical measures for degree correlation.* One of the widely used measures for degree correlation is ANND [3],

$$k_{nn}(k) = \sum_j j p_c(j|k), \qquad (1)$$

where $p_c(j|k) = e_{jk} / \sum_j e_{jk} = e_{jk}/q_k$ is the conditional probability that an edge belonging to a node with degree $k$ will connect to a node with degree $j$. This measure considers the average degree of the neighbors of a node as a function of its degree $k$, and it can provide a clear indication for the presence or absence of degree correlation in networks. When it is independent of $k$, meaning that networks have no obvious correlation of degree. In homogeneous and uncorrelated networks, $k_{nn}(k) = \langle k \rangle$, while it will increase with the increase of the heterogeneity of networks: $k_{nn}(k) = \langle k^2 \rangle / \langle k \rangle$ ( $\langle k \rangle = \sum_k k p_k$ and $\langle k^2 \rangle = \sum_k k^2 p_k$ ) in general uncorrelated networks. In general correlated networks, ANND will increases with $k$ for assortative mixing, meaning that nodes preferentially attach to other nodes with similar degrees, while it will decrease with $k$ for dissassortative mixing, meaning that high-degree nodes preferentially attach to other low-degree nodes. The representation above provides a plain interpretation for the origin of degree correlations. However, this quantity can give a clear but only qualitative characterization for the degree correlation in networks.

A more coarse-grained and quantitative characterization for degree correlation in networks can be given by DCC [2],

$$r = \frac{1}{\sigma_q^2} \sum_{jk} jk(e_{jk} - q_j q_k), \qquad (2)$$

where $\sigma_q^2 = \sum_{jk} k^2 q_k - (\sum_k k q_k)^2$. It can describe the level of degree correlation in networks by a scalar quantity ( $-1 \le r \le 1$ ). Compared with the ANND's qualitative description, DCC can not only give the tendency of the degree correlation by its sign, but it can also give a value being able to reflect the strength of the correlation. So it is very convenient to analyze and compare the degree correlation of different networks by DCC.

*Analysis of statistical measures for degree correlation.* The definitions of both ANND and DCC depend on the degree-degree joint probability. So the two measures must be closely related. The quantitative relation between the two measures can be given by the following transformation (from the equation (2)),

$$r = \frac{1}{\sigma_q^2} [\sum_k k q_k \sum_j j e_{jk} / q_k - \sum_k k q_k \sum_j j q_j]$$

$$= \frac{1}{\sigma_q^2} [\sum_k k q_k k_{nn}(k) - \sum_k k q_k \sum_j q_j k_{nn}(j)]$$

$$= \frac{1}{\sigma_q^2} \sum_{jk} q_k k q_j [k_{nn}(k) - k_{nn}(j)]$$

$$= \frac{1}{\sigma_q^2} \sum_{j,k(j<k)} \{q_k k q_j [k_{nn}(k) - k_{nn}(j)] + q_j j q_k [k_{nn}(j) - k_{nn}(k)]\}$$

$$= \frac{\sum_{j,k(j<k)} q_k q_j (k-j)[k_{nn}(k) - k_{nn}(j)]}{\sum_{j,k(j<k)} q_k q_j (k-j)^2},$$

(3)

where,

$$\sigma_q^2 = \sum_{jk} k^2 q_k - (\sum_k k q_k)^2 = \sum_{jk} k q_k q_j (k-j)$$

$$= \sum_{j,k(j<k)} [k q_k q_j (k-j) + j q_j q_k (j-k)] \qquad , \quad (4)$$

$$= \sum_{j,k(j<k)} q_k q_j (k-j)^2$$

and the constraint condition due to the topological constraint of the network is used,

$$\sum_k q_k k_{nn}(k) = \sum_k q_k k = \frac{\langle k^2 \rangle}{\langle k \rangle}. \qquad (5)$$

(Note that it is related to the sum of the degrees of ends of all edges in network). According to the equation (3), $r=0$, when $k_{nn}(k) \equiv C$, meaning the absence of the degree correlation. $\because k \ge j$, $\therefore q_k q_j (k-j) \ge 0$. In assortative networks, $k_{nn}(k)$ is a monotonically increasing function with $k$, so $k_{nn}(k) - k_{nn}(j) > 0$ (when $k \ge j$). As a result, $r$ will be a positive value in assortative networks, and thus the assortative mixing of degree is also called as *positive* correlation. In disassortative networks, a monotonically decreasing $k_{nn}(k)$ corresponds to a negative $r$-value, so the disassortative mixing of degree is often called *negative* correlation. Clearly, the two measures for describing the degree correlation are absolutely consistent.

*Linear relation in degree correlation.* As we know, ANND as a function of $k$ gives a curve that varies with $k$. It can be characterized by suitable fitting functions. For example, researchers showed a power-law dependence of ANND on degree [3, 38]. Ma and Szeta extended the Aboav-Wearie law to the analysis of degree correlation in complex networks [25]. In the study of complex systems, the linear analysis is often more appreciated, due to the simplicity and clarity of it. Here, we show that the results of interest can be obtained by using the linear fitting function ( $ak+b$ ). Substituting $k_{nn}(k)$ in the equation (3) by the linear function, and using the constraint equation (5), we can obtain,

$$a = r \quad \text{and} \quad b = \frac{\langle k^2 \rangle}{\langle k \rangle}(1-r). \qquad (6)$$

Clearly, the slope $a$ in the linear relation corresponds to the correlation coefficient $r$. The results suggest that DCC can not only describe the level of the degree correlation in networks, but can also reflect the speed that ANND varies with $k$.

Moreover, for a perfect assortative network, $r=1$, while ANND will be a perfect linear function $k_{nn}(k)=k$. This means that only nodes with the same degree are connected with each other. For disassortative networks, ANND is a descreasing function of $k$. Particularly, for perfect negative correlation networks, $r=-1$ and then $k_{nn}(k) = -k + 2\langle k^2\rangle/\langle k\rangle$. In uncorrelated networks, $r=0$, while $k_{nn}(k) = \langle k^2\rangle/\langle k\rangle$.

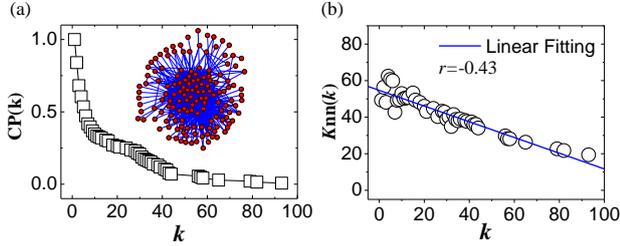

**FIG. 1:** (Color Online) (a) The airline network of China (ANC) and the cumulative degree distribution CP(k) of the network. The (undirected and un-weighted) network describes the air routes among all cities with operating airports in mainland China (excluding Taiwan, Hong Kong and Macao) from October 28, 2007 to March 29, 2008 . (b) The data of ANND as a function of $k$, and DCC in the network. The blue solid line is the linear fitting of ANND in the network.

*Application to the airline network of China.* There exist some real-world networks where ANND may be more suitably characterized by the linear relation. In Fig. 1 (a), we show the airline network of China (ANC) [39] where the cities are denoted by nodes and the air routes are denoted by links, and the cumulative degree distribution CP (k) of the network. In the network, a few busy cities have a large number of air routes, dominating the transportation system, the number of routes of each city decreases quickly, and most of small cities have only 1-3 air routes. In Fig. 1 (b), we show the data of ANND as a function of $k$ and DCC in ANC. Clearly, ANND is a decreasing function of $k$ as a whole, and DCC is also negative. That is to say, the characterization of ANND and DCC for degree correlation is consistent. In Fig. 1 (b), the linear fit of ANND is obviously more suitable for the network than others such as the power-low fit. Fitting the data of ANND by a linear function, we find that the slope of the linear fit exactly corresponds to DCC in the network.

**Effect of structural variation on degree correlation.** –First, as an example, the effect of structural change on the degree correlation is displayed in a simple and visual network, caused by the addition of one node and one edge, and then we carefully investigate the perturbation of the degree correlation in general complex networks, caused by the "rich club".

*Effect of addition of node and edge.* In terms of the linear relation between ANND and DCC, DCC can characterize the level of degree correlation in whole networks, but can also reflect the monotonicity of ANND and the speed that ANND varies with $k$ as a whole. Inversely, by ANND, one can also learn about the type of the correlations and the strength of such correlations. In Ref [40], Estrada shows that the assortativity of networks can be affected by simple structural changes, and gives an interesting explanation for the phenomena. In Fig. 2, Net A is disassortative, while Net B becomes assortative, though the two networks have only simple structural difference caused by the addition of one node and one edge. We can see that ANND in Net A is a decreasing function of $k$, while ANND in Net B becomes an increasing function of $k$ on the whole. Naturally, $r<0$ for Net A, while $r>0$ for Net B, according to the linear relation between $r$ and ANND. Moreover, we also notice that both the statistical measures for degree correlation seem to be too sensitive in such (quasi-)homogeneous networks.

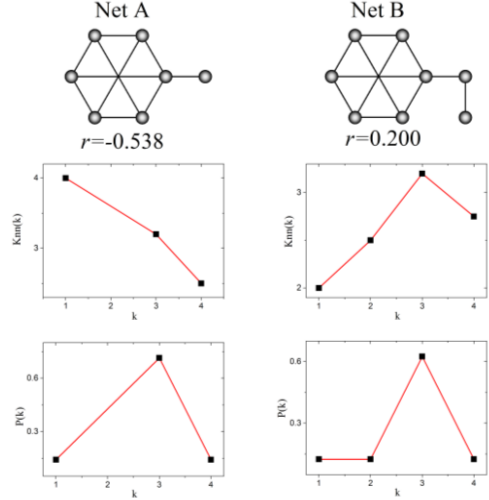

**FIG. 2:** (Left) Network (Net A) constructed by linking a regular graph and a single node, and ANND and the degree distribution P(k) of the network. This network is disassortative in terms of $r$ and/or ANND. (Right) Network (Net B) which has more nodes and edges than Net A by adding one node and one edge, and ANND and P(k) of the network. This network is assortative in terms of $r$ and/or ANND.

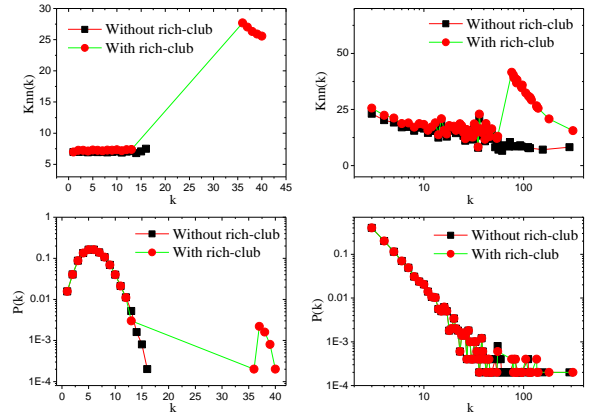

**FIG. 3:** (Left) For Erdös-Rényi (ER) network with 5000 nodes and 30000 edges [41], we show the ANND and the degree distribution P($k$) in the homogeneous network with "rich club" (0.5% nodes) and without "rich club". The original ER network is uncorrelated in degree in terms of $r$ and/or ANND. (Right) For BA scale-free network where the number of nodes and the mean degree are the same as in the ER networks [42], we also show ANND and P($k$) in the heterogeneous network with "rich club" (0.5% nodes) and without "rich club".

*Effect of Rich club.* Here, we choose the top 0.5% of nodes with the highest degrees as the set of rich nodes in a network and manipulate the connections among the rich nodes: (1) remove the edges among the rich nodes, so there is no rich-club in the network; (2) make the rich nodes fully connected to each other, so they form a fully connected sub-graph. The main topological structures in the networks are the same for the two types of networks above except for the connection pattern among the rich nodes. The rich club can control the assortativity in some networks [43], but what determines the contrability of the rich club for the assortativity in the networks? Fig. 3 shows two typical examples of the rich club affecting the degree correlation. The rich club can change more strongly ANND in the ER network, leading to the large change of the slope (i.e. DCC) of the linear fit of it. From the absence to presence of the "rich club", DCC in the ER network changes from 0.00 to 0.53, while, for the BA scale-free network, it only changes from -0.08 to 0.04. Clearly, the "rich club" can affect more strongly the assortativity of the ER network.

In the above networks, there is the same number of nodes and the same mean degree, then what does lead to the difference? In fact, the two networks above have very different topological features, especially the degree distribution. The "rich club" can more strongly affect the degree distribution of the ER network (see Fig. 3). The degree distribution is very narrow for the ER network, while it is very broad for the BA scale-free network, following the power-law distribution. The range of the degrees in a network is closely related to the contrability of the rich club for the degree correlation, because, in terms of the linear analysis of ANND, the larger the range of degrees in the networks is, the more difficultly the rich club changes the slope of ANND (i.e. DCC). This also suggest that the (quasi-) homogeneous networks (with narrow degree distribution) such as ER are very sensitive to "rich club", while the heterogeneous networks (with broad degree distribution) such as BA are insensitive.

**Table 1:** Nine undirected networks: SW denotes the network generated by the small-world model, ER denotes the network generated by Erdös-Rényi model, PG denotes the U.S. Power Grid, COND denotes the scientific collaboration networks on condensed matter physics, EPA denotes the network from the pages linking to www.epa.gov, BA denotes the network generated by the scale-free model, AS denotes the Internet topology at the level of autonomous systems, PFP denotes the network generated by the model for the Internet topology and BOOK denotes the word adjacency network from Darwin's "The Origin of Species". The proportion of rich nodes in the networks is 0.5% except COND (0.2% nodes as rich nodes because it has larger scale than other networks). $N$ is the number of nodes in the networks, $\langle k \rangle$ is the mean degree of the networks, $k_{max}$ is the maximal degree; $r$ and $r'$ denote the values of DCC in the networks without "rich-club" and with "rich-club"; $\Delta r = r - r'$ denotes the increase of DCC; $k_R$ is the mean degrees of the rich nodes in the networks with "rich-club".

| Network | $N$ | $\langle k \rangle$ | $k_{max}$ | $r$ | $r'$ | $\Delta r$ | $k_R$ |
|---|---|---|---|---|---|---|---|
| SW | 5000 | 6.0 | 15.5 | 0.00 | 0.65 | 0.65 | 34.1 |
| PG | 4941 | 2.7 | 19.0 | -0.01 | 0.60 | 0.61 | 34.2 |
| ER | 5000 | 6.0 | 16.0 | 0.00 | 0.53 | 0.53 | 38.7 |
| Cond | 16726 | 5.7 | 107.0 | 0.17 | 0.32 | 0.15 | 94.7 |
| EPA | 4772 | 3.7 | 175.0 | -0.31 | -0.15 | 0.16 | 128.4 |
| BA | 5000 | 6.0 | 218.6 | -0.08 | 0.04 | 0.12 | 150.4 |
| AS | 5375 | 3.9 | 1193 | -0.19 | -0.19 | 0.00 | 216.6 |
| PFP | 5000 | 6 | 1258.8 | -0.25 | -0.24 | 0.01 | 354.0 |
| BOOK | 7724 | 11.4 | 2568 | -0.24 | -0.24 | 0.00 | 628.5 |

In Table I, we further list the results of 9 different un-directed networks: 5 real-world networks and 4 model networks, arranged with $\Delta r$ increasing. $\Delta r$ is closely related to many factors such as the mean degree, the size and the heterogeneity of the networks and so on. $k_{max}$ denotes the upper bounds of degrees in the networks, which can largely reflect the range of degrees in the networks. As expected, on the whole, the smaller $k_{max}$, the larger $\Delta r$. Based on $\Delta r$ or $k_{max}$, the networks can be classified into three different groups. In the first group (SW, PG and ER) with small values of $k_{max}$ (<100), the rich club strongly changes DCC. In the second group (EPA, Cond and BA) with $100<k_{max}<1000$, the rich club has slight effect on DCC. In the third group (AS, PFP and BOOK) with large values of $k_{max}$ (>1000), the rich club hardly changes DCC in the networks.

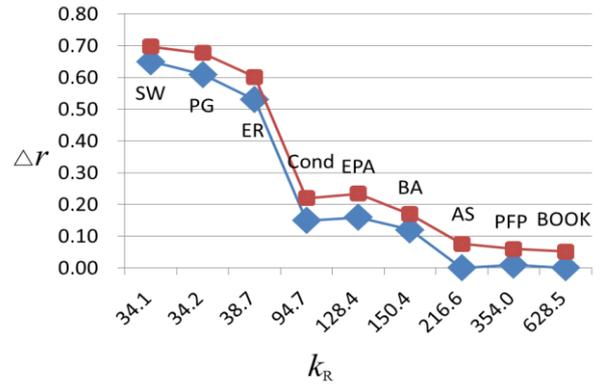

**FIG. 4:** The real values (◆) and estimated values (■) of the increase $\Delta r$ of DCC caused by the rich club in different networks.

The rich club manipulates the connections of a small proportion of rich nodes, affecting the mean degrees of the rich nodes ($k_R$). $k_R$ is also crucial to the changes of DCC. On the whole, the larger $k_R$ corresponds to the smaller $\Delta r$ (see Table I and Fig. 4). In terms of the linear analysis of degree correlation, we can give a simple estimation of $\Delta r$ caused by the rich club. The mean degrees of the nearest neighbors of most nodes and the rich nodes in the networks can be estimated by $k_{nn}(\langle k \rangle) = b + r \cdot \langle k \rangle$ and,

$$k_{nn}(k_R) = \frac{(b + r \cdot k_r)k_r + (N_r - 1)k_R}{k_R}, \quad (7)$$

where $b$ is the intercept in equation (7), $r$ is the degree correlation in the original network, $\langle k \rangle$ is the mean degree of the whole networks, $N_r$ is the number of the rich nodes, while $k_r$ and $k_R$ are the mean degrees of the rich nodes in the networks without rich club and with rich club ($k_R \approx k_r + N_r - 1$). Then the estimation of $\Delta r$ can be written as

$$\Delta r = \frac{k_{nn}(k_R) - k_{nn}(\langle k \rangle)}{k_R - \langle k \rangle} - r. \quad (8)$$

Fig. 4 shows that the overall trend of the estimation of $\Delta r$ is consistent with the real values of it, though this is only a simple linear evaluation for the degree correlation.

**Conclusion and discussion.** - We analyzed the statistical measures for characterizing the degree correlation in networks, gave the exact proof of the consistency of the measures, and then exhibited the general linear relation in the degree correlation, which provides a simple and interesting perspective on the analysis of the degree correlation. By comparing the other analysis methods [3, 25, 38] which are difficult to explicitly give the analytical relations of the fitting parameters, we showed that the slope of the linear fit of ANND corresponds exactly to DCC, meaning that DCC can not only characterize the strength of the degree correlation in network, but can also reflect the speed that ANND varies with $k$. And then, as an exemplification for the results above, we analyzed the linear degree correlation in the airline network of China.

In some cases, the assortativity of networks can be affected easily by small structural changes. In terms of the general linear relation in the degree correlation, we analyze the perturbation of the degree correlation in networks caused by the addition of few nodes and the "rich club". The slope of ANND in homogeneous networks such as the ER graphs is affected strongly by the simple structural changes, while it has only a relatively slight variation for heterogeneous networks such as the BA networks. According to the linear relation between $r$ and ANND, DCC in homogeneous networks is naturally sensitive much more than in heterogeneous networks with broad degree distribution. It may not be good news that the statistical measures seem too sensitive in homogeneous networks.

To our knowledge, the concept of degree correlation is used to reveal the statistical feature in heterogeneous networks. Whether it should be applied to (quasi-)homogeneous networks (which have narrow degree distribution). How to understand the results that such statistical quantities as DCC and ANND generate in the networks? The stability of the measures is related to heterogeneity of networks. When one applies these statistical measures to the analysis of the degree correlation in complex networks, maybe the heterogeneity of networks should also be an important factor to be considered. We will discuss this topic in depth in future work. Moreover, it is reported that DCC has the size dependence [37]. Thus the linear analysis, due to the close connection with DCC, is also size dependent, while it provides a complementary method for the network analysis. Finally, we hope that the work can help to further understand the property of the degree correlation in networks.

**Acknowledgement.** - This work has been supported by the Scientific Research Fund of Education Department of Hunan Province (Grant Nos. 14C0126, 11B128, and 14C0127), and Natural Science Foundation of Hunan Province (Grant No. 13JJ4045), and partly by the Doctor Startup Project of Xiangtan University (Grant No. 10QDZ20).